\documentclass[twocolumn,prl,superscriptaddress]{revtex4}
\usepackage{amsfonts}
\usepackage{amssymb}
\usepackage{amsmath}
\usepackage{epsfig}
\usepackage{color}
\usepackage{graphics, graphicx}
\usepackage{bbold}
\usepackage{psfrag}
\usepackage{mathcomp}
\usepackage{subfigure}
\usepackage{verbatim}
\usepackage[colorlinks,citecolor=blue]{hyperref}

\makeatletter

\newcommand{\Rmnum}[1]{\expandafter\@slowromancap\romannumeral #1@}
\makeatother

\begin{document}

\title{Observation of Large-Number Corner Modes in $\mathbb{Z}$-class Higher-Order \\Topolectrical Circuits}
	
\author{Yi Li}
	\affiliation{State Key Laboratory of Quantum Optics and Quantum Optics Devices, Institute
		of Laser Spectroscopy, Shanxi University, Taiyuan, Shanxi 030006, China}
	\affiliation{Collaborative Innovation Center of Extreme Optics, Shanxi
		University,Taiyuan, Shanxi 030006, China}
	
\author{Jia-Hui Zhang}
	\affiliation{State Key Laboratory of Quantum Optics and Quantum Optics Devices, Institute
		of Laser Spectroscopy, Shanxi University, Taiyuan, Shanxi 030006, China}
	\affiliation{Collaborative Innovation Center of Extreme Optics, Shanxi
		University,Taiyuan, Shanxi 030006, China}
	
\author{Feng Mei}
	\email{meifeng@sxu.edu.cn}
	\affiliation{State Key Laboratory of Quantum Optics and Quantum Optics Devices, Institute
		of Laser Spectroscopy, Shanxi University, Taiyuan, Shanxi 030006, China}
	\affiliation{Collaborative Innovation Center of Extreme Optics, Shanxi University,Taiyuan, Shanxi 030006, China}
	
\author{Biye Xie}
	\email{xiebiye@cuhk.edu.cn}
	\affiliation{School of Science and Engineering, The Chinese University of Hong Kong, Shenzhen, Guangdong 518172, China}

\author{Ming-Hui Lu}
\email{luminghui@nju.edu.cn}
\affiliation{College of Engineering and Applied Sciences and National Laboratory of Solid State Microstructures, Nanjing University,	Nanjing 210093, China}
\affiliation{National Laboratory of Solid State Microstructures, Collaborative Innovation Center of Advanced Microstructures, Nanjing University, Nanjing, 518172, China.}
	
\author{Jie Ma}
	\email{mj@sxu.edu.cn}
	\affiliation{State Key Laboratory of Quantum Optics and Quantum Optics Devices, Institute
		of Laser Spectroscopy, Shanxi University, Taiyuan, Shanxi 030006, China}
	\affiliation{Collaborative Innovation Center of Extreme Optics, Shanxi
		University,Taiyuan, Shanxi 030006, China}
	
\author{Liantuan Xiao}
	\affiliation{State Key Laboratory of Quantum Optics and Quantum Optics Devices, Institute
		of Laser Spectroscopy, Shanxi University, Taiyuan, Shanxi 030006, China}
	\affiliation{Collaborative Innovation Center of Extreme Optics, Shanxi
		University,Taiyuan, Shanxi 030006, China}

\author{Suotang Jia}
	\affiliation{State Key Laboratory of Quantum Optics and Quantum Optics Devices, Institute
		of Laser Spectroscopy, Shanxi University, Taiyuan, Shanxi 030006, China}
	\affiliation{Collaborative Innovation Center of Extreme Optics, Shanxi
		University,Taiyuan, Shanxi 030006, China}
	
\date{\today }
	
\begin{abstract}
  Topological corner states are exotic topological boundary states that are bounded to zero-dimensional geometry even the dimension of bulk systems is large than one. As an elegant physical correspondence, their numbers are dictated by the bulk topological invariants. So far, all previous realizations of HOTIs are hallmarked by $\mathbb{Z}_2$ topological invariants and therefore have only one corner state at each corner. Here we report an experimental demonstration of $\mathbb{Z}$-class HOTI phases in electric circuits, hosting $N$ corner modes at each single corner structure. By measuring the impedance spectra and distributions, we clearly demonstrate the $\mathbb{Z}$-class HOTI phases, including the zero-energy corner modes and their density distributions. Moreover, we reveal that the local density of states (LDOS) at each corner for $N=4$ are equally distributed at four corner unit cells, prominently differing from $\mathbb{Z}_2$-class case where the LDOS only dominates over one corner unit cell. Our results extend the observation of HOTIs from $\mathbb{Z}_2$ class to $\mathbb{Z}$ class and the coexistence of spatially overlapped large number of corner modes which may enable exotic topological devices that require high degeneracy boundary states.
\end{abstract}

\maketitle

{\textit{Introduction}. Bulk-boundary correspondence, as one of the most important physical properties of topological materials, links the bulk topological invariant to the number of boundary states at a certain open boundary~\cite{Kane2010,Zhang2011}. For $\mathbb{Z}$-class topological phases, the topological invariant is no longer limited to unity and can be a larger number, consequently generating large numbers of topological boundary states~\cite{Kane2010,Zhang2011}. Besides the fundamental interests, multiple topological boundary states can significantly improve the channel capacities and coupling efficiencies, having great practical value in enabling functional topological photonic, acoustic and mechanical devices~\cite{Lu2014rev,Huber2016rev,Khan2017rev,Lu2018rev,Lu2019rev,Chan2019rev,Zhang2022rev,Alu2022rev}. Therefore, realizing such unusual topological phase is highly desirable but very hard, until recently the topological insulator phases with larger chern numbers~\cite{Wang2013,Fang2014,Marin2014,An2016} have been experimentally reported in photonic crystals~\cite{Marin2015} and solid materials~\cite{Zhao2020}.

\begin{figure*}
 	\includegraphics[width=0.8\textwidth,height=0.4\textwidth]{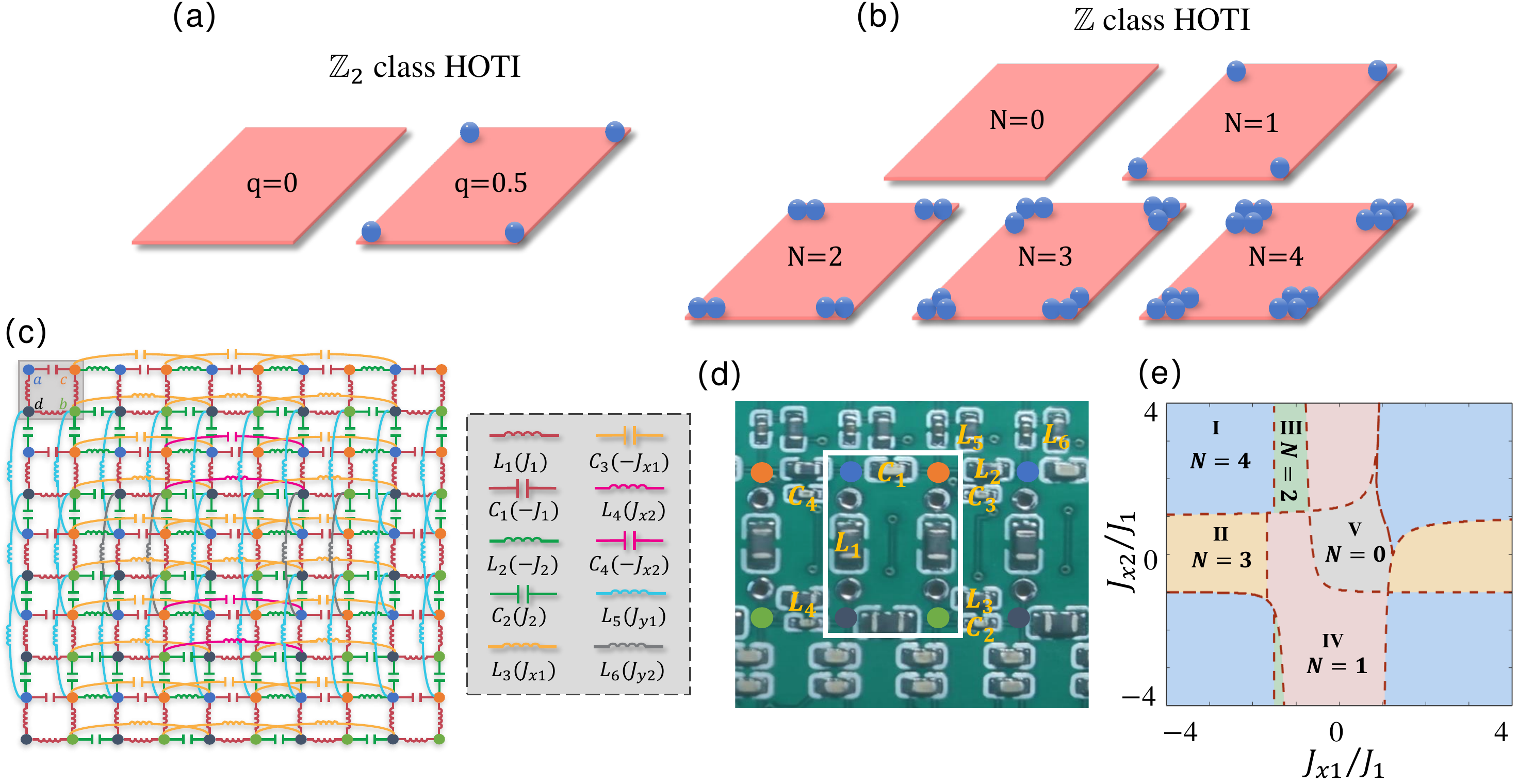}
 	\caption{(a) $\mathbb{Z}_2$-class HOTI phases, characterized by quantized quadrupole moments, featuring one corner mode at each corner. (b) $\mathbb{Z}$-class HOTI phases, protected by multipole chiral numbers, hosting multiple corner modes in each corner. (c) Circuit diagram for implementing a long-range coupling lattice model supporting $\mathbb{Z}$-class HOTI phase. Grey square lists the specific circuit elements and the corresponding lattice parameters. (d) Experimentally realized circuit for a unit cell. (e) Topological phase diagram in the parameter space of $J_{x1}$ and $J_{x2}$ for $J_{y1}=J_{x1}$, $J_{y2}=J_{x2}$ and $J_2=-0.5J_1$. }
 	\label{Fig1}
 \end{figure*}

The recent discovery of higher-order topological insulator (HOTI) phases has further expanded the concept of topological phases~\cite{HOTI2017a,HOTI2017b,HOTI2017c,HOTI2018}. One appealing feature associated with such phase is the existence of lower-dimensional topological boundary states, for example, the zero-dimension topological corner states. Topological corner states have recently attracted great interests in many platforms~\cite{Xie2021rev,Jacob2020rev}, such as photonics~\cite{Bena2018a,Bena2018b,Ashraf2019,Lu2019,Dong2019,Hafezi2019,Lu2020,Wang2021,Freqymann2022,Chen2022}, phononics~\cite{Serra2018,Xue2019,Khan2019,Jiang2019a,Jiang2019b,Zhang2019,Khan2020,Liu2020,Qiu2021,Jiang2022a,
Jiang2022b,Junkai2022,Liu2021} and electric circuits~\cite{Imhof2018,Ezawa2018,ZhangX2019,LiuZ2020,Yan2020a,Yan2020b,ZhangS2020,ZhangX2020,ZhangX2021,Xu2021,Hughes2022,Gorlach2022,ZhangX2022a,ZhangX2022b,Yan2022}. They have also motivated numerous important applications and studies, ranging from topological cavities and lasers~\cite{Park2020,Xu2020a} to nonlinear~\cite{Fleury2019,Heinrich2021,Chen2021,Ezawa2022} and quantum optics~\cite{Xu2020b,Li2022}. The latest finding is that HOTIs are not limited to $\mathbb{Z}_2$ class, which can be further promoted to $\mathbb{Z}$ class~\cite{Bena2022}. As illustrated in Figs. \ref{Fig1}(a,b), distinct from $\mathbb{Z}_2$ class HOTI phases, $\mathbb{Z}$-class HOTI phases are classified by multiple chiral numbers, featuring large-number topological corner modes at each corner. However, their model realizations require long-range couplings, which render the experimental demonstration of such unconventional phases extremely challenging.

In this Letter, we report the experimental realization and detection of $\mathbb{Z}$-class HOTI phases using topolectrical circuits. Electric circuits recently have been demonstrated being a powerful platform for exploring topological phases~\cite{Lee2018,Zhao2018,Roy2021}. In this work, the electric circuits are designed and fabricated with dimerized long-range couplings. Based on calculating multiple chiral numbers $N$ and circuit admittance spectra, we demonstrate that our constructed circuits support four distinct nontrivial $\mathbb{Z}$-class HOTI phases identified by $N=1,2,3,4$ and host $N$ corner modes at each corner. Experimentally, the measured zero-energy corner-mode resonances and distributions agree well with the simulated results, demonstrating the emergence of $\mathbb{Z}$-class HOTI phases in the circuits. Moreover, in our topolectrical circuit experiments we implement LDOS measurements. Via detecting LDOS, we successfully measure the multiple chiral number $N$, which is an inalienable prediction of $\mathbb{Z}$-calss topological classification and allows us to experimentally distinguish different $\mathbb{Z}$-class HOTI phases. We also study the fractional corner charges associated with $\mathbb{Z}$-class HOTI phases. Different from $\mathbb{Z}_2$-class HOTI phases that have a single $\frac{1}{2}$ fractional charge at each corner, we find that $\mathbb{Z}$-class HOTI phases have $N$ such fractional charges neatly distributed in each corner.

In Fig. \ref{Fig1}(c), we design an electric circuit to demonstrate its power in building long-range couplings, which are crucial and valuable for realizing and exploring various complex topological phases. The response of a circuit at the frequency $\omega$ is given by the Kirchhoff's law $I_a=\sum_bJ_{ab}(\omega)V_b(\omega)$, where $I_a$ is the input current flowing out of node $a$, $V_b(\omega)$ is the voltage of the circuit node $b$ and $J_{ab}(\omega)$ is the element of circuit Laplacian. At the resonant frequency, the circuit Laplacian function as the analogue of lattice model Hamiltonian, where the circuit nodes are referred to as the lattice sites and the connections between different nodes through inductors or capacitors are corresponding to the lattice couplings.  Specifically for Fig. \ref{Fig1}(c), each unit cell contains four nodes labelled by $a-d$, as exhibited in Fig. \ref{Fig1}(d) for its realization on the printed circuit board. The long-range couplings in our work are chosen between next-nearest-neighbor unit cells, but which can be easily generalized to much longer. The positive and negative couplings are respectively implemented through inductors and capacitors. Different from the original $\mathbb{Z}$-class HOTI model~\cite{Bena2022}, we find that as the long-range intra-cell couplings are designed to feature dimerized configurations in the even rows and columns, the corresponding circuit has much richer topological phase transitions, as presented in Fig. \ref{Fig1}(e).

\begin{figure*}
 	\includegraphics[width=0.9\textwidth,height=0.55\textwidth]{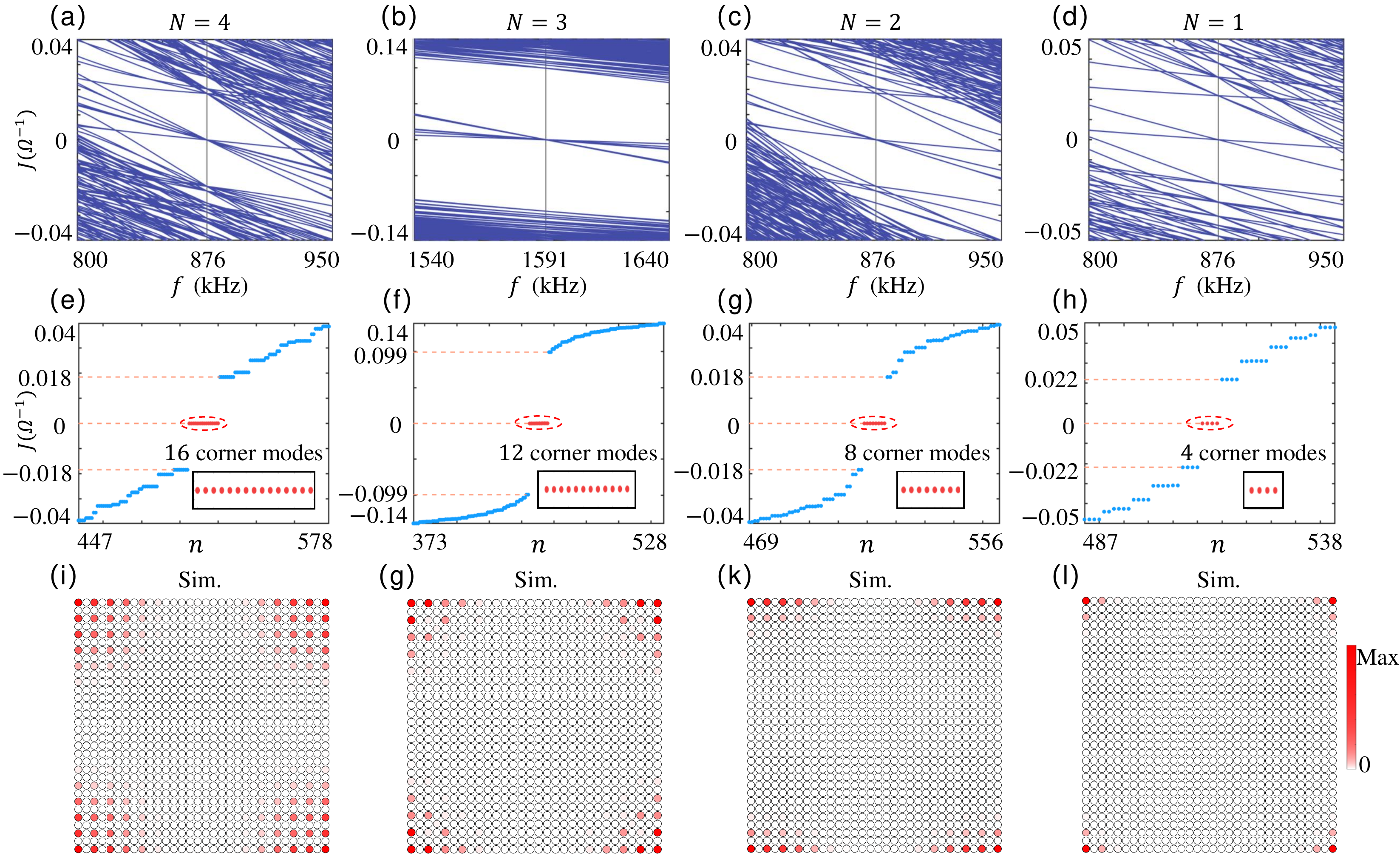}
 	\caption{(a-d) Simulated circuit Laplacian (admittance) spectra as a function of driving frequency for different MCNs. At the resonant frequencies (gray solid lines), the spectra respect chiral symmetry (red dashed lines) and host large-number ($4N$) zero modes that are corresponding to corner modes, as manifested by sorting them out in (e-h). (i-l) show the distributions of the zero-energy corner modes, unveiling that the corner modes in $\mathbb{Z}$-class HOTI phases have extended mode areas.}
 	\label{Fig2}
 \end{figure*}

The long-range couplings in Fig. \ref{Fig1}(c) are designed in a chiral-symmetry fashion. Thus, the circuit lattice naturally inherits chiral symmetry, and its $\mathbb{Z}$-class higher-order bulk topology is characterized by the multiple chiral number (MCN), which is a real-space topological invariant defined as
\begin{equation}
 {N}=\frac{1}{{2\pi i}}\text{Tr}\log (\bar Q_{xy}^A\bar Q_{xy}^{B\dag}),
\end{equation}
where $\bar{Q}_{xy}^{A,B}$ are the multipole moment operators projected into the sublattice spaces. By numerically calculating the MCN, the topological phase diagram associated with the implemented lattice model Hamiltonian in Fig. \ref{Fig1}(c) is presented in Fig. \ref{Fig1}(e). The result shows that the presence of long-range couplings could drive the circuit into four distinct nontrivial $\mathbb{Z}$-class HOTI phases, as indicated by $N=1,2,3,4$. Particularly, compared with the original work~\cite{Bena2022}, implementing the $\mathbb{Z}$-class HOTI phase $N=3$ does not require complicated diagonal long-range couplings. Moreover, for the nontrivial cases $N$\textgreater$1$, the corresponding system features trivial quadrupole moment $q=0$. However, as shown below, they indeed host large-number topological corner modes, satisfying bulk-corner correspondences.

To demonstrate the emergence of multiple corner modes in the circuits, in Figs. \ref{Fig2}(a-d) we simulate the eigenvalues of the circuit Laplacians varying with the driving frequency, for the circuits featuring different MCNs. Grounding conditions (see Supplementary Materials for the detailed grounding circuits) are taken into account in the simulations so that all diagonal terms in the circuit Laplacian vanish at the resonant frequencies. As manifested, the corresponding spectra respect chiral symmetry at the resonant frequencies $f_0=876$ kHz for $N=1,2,4$ and $f_0=1591$ kHz for $N=3$. By sorting them out in Figs. \ref{Fig2}(e-h), we can see that the circuit Laplacians support large numbers of gapped zero modes, with their specific numbers given by $4N$, determined by the MCNs. Figs. \ref{Fig2}(i-l) plot the corresponding distributions in the circuit lattices for these zero modes, as indicated which are the corner modes, together with Figs. \ref{Fig2}(e-h) confirming $\mathbb{Z}$-class higher-order bulk-corner correspondences. Compared to $\mathbb{Z}_2$-class HOTI phases, the corner modes in $\mathbb{Z}$-class HOTI phases have much larger mode areas and richer distribution features. For example, Figs. \ref{Fig2}(i-l) show that as $N$ increases from 1 to 4, the corner mode area could increase from 1 unit cell to roughly 16 unit cells, which may can motivate novel applications in designing topological corner-mode lasers.

\begin{figure*}[htbp]
	\includegraphics[width=0.9\textwidth,height=0.6\textwidth]{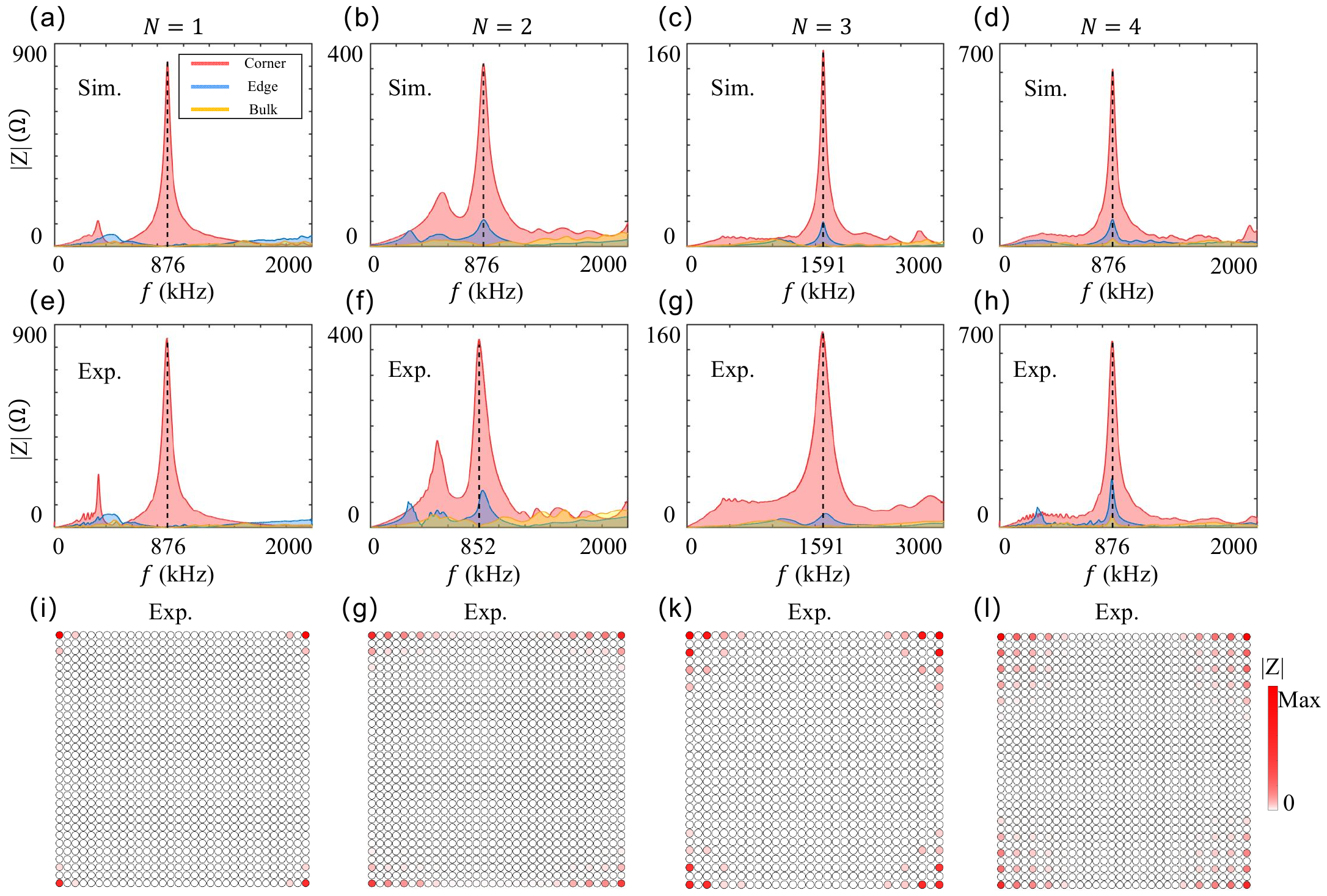}
	\caption{(a-d) Simulated and (e-h) measured impedances between the corner, edge and bulk circuit nodes and the ground via scanning the driving frequency in different $\mathbb{Z}$-class higher-order topological circuits. Both results clearly verify that the corner mode resonances occur at the resonant frequencies (black dashed lines). \textbf{i-l} Measured impedance distributions for the corner modes, demonstrating that the mode areas in $\mathbb{Z}$-class HOTI phases scales up with the MCNs.}
	\label{Fig3}
\end{figure*}

We experimentally fabricate four nontrivial $\mathbb{Z}$-class higher-order topolectrical circuits, containing $16\times16$ unit cells, corresponding to $N=1,2,3,4$. Their featured higher-order topological features, i.e., the zero-energy corner-mode resonances and localizations (as predicted in Fig. \ref{Fig2}), are detected by measuring the impedance $Z_{a}$ between the node $a$ and the ground, which is related to the eigenvalues and eigenvectors of the circuit Laplacian, $j_n$ and $\psi_n$ respectively, via $Z_{a}=\sum_{n}|\psi_{n,a}|^2/j_n$. This relationship enables that, when the circuit is excited at the resonant frequency, the impedances at corner nodes should be extremely large compared to impedances at the bulk and edge nodes, as in the corner nodes only zero-energy corner modes have been mostly excited, which leads to that $|\psi_{n,a}|^2$ is maximal and $j_n$ is near zero at corners. Figs. \ref{Fig3}(a-d) verify this feature for four different MCNs based on simulating the measurements using the simulation software LTspice. As expected, the highest peaks in the simulated corner impedances occur at the resonant frequencies, and the small peaks in the edge impedances reflect that the distribution of corner modes extend to edge nodes in the $\mathbb{Z}$-class HOTI phases with larger MCNs. The experimentally measured corner, edge and bulk impedances are presented in Figs. \ref{Fig3}(e-h) respectively for different MCNs, which agree very well with the simulation results. The corresponding distributions in the circuits for the corner modes are probed by measuring all-node impedances at the resonant frequencies. As displayed in Figs. \ref{Fig3}(i-l), as the topological invariant $N$ increases from $1$ to $4$, the corner mode areas will extend to the edge and bulk nodes, scaling up with the MCNs, in accordance with theoretical simulations in Figs. \ref{Fig2}(i-l).

\begin{figure*}[htbp]
   	\includegraphics[width=1 \textwidth,height=0.5\textwidth]{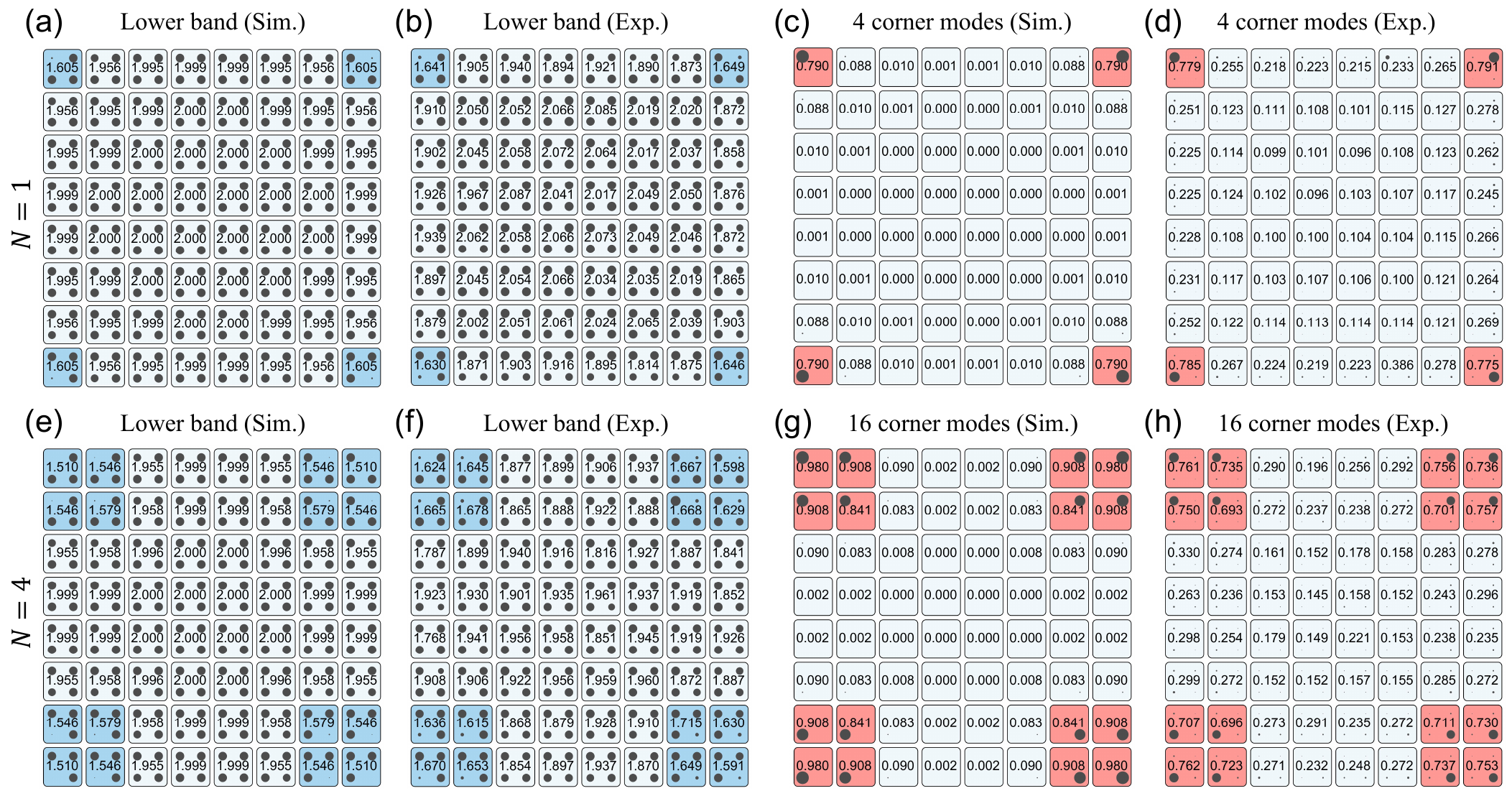}	
   	\caption{Spatial maps of simulated and measured LDOS for $N=1$ and $N=4$ integrating over (a,b,e,f) the lower band, and (c,d,g,h) an in-gap spectral range covering all corner modes. Each circuit node is represented as a circle with radius proportional to the LDOS. Each number gives the LDOS in the corresponding unit cell. The topological invariants $N$ and fractional corner charges are respectively detected and manifested by the LODS in the shaded corner unit cells.}
   	\label{Fig4}
\end{figure*}

To unambiguously distinguish different $\mathbb{Z}$-class HOTI phases, it is highly desirable to have the capability to detect the MCN values. The simplest way is to count the numbers of corner modes which equal the MCNs according to bulk-corner correspondence. However, the corner modes are degenerate in the admittance spectra and can not be extracted out. Now we exhibit that such topological numbers can be probed by LDOS measurement. Notice that using LDOS measurements to probe fractional charges have recently attracted great interests in photonic and acoustic systems~\cite{BahlG2020,Dis2021a,Dis2021b,JiangJH2022,ZhangS2022b,GeH2023}. For circuit systems, the LDOS (mode density) at each circuit node, for example the circuit node $a$, is measured via the real part of the impedance $Z_a$, i.e., $\rho(f,a)=2f \,\text{Re}[Z_a]$, see Supplementary Materials for all measured LDOS data and more details. In Fig. \ref{Fig4}, we exemplify the $\mathbb{Z}$-class topological phases $N=1$ and $N=4$. To extract the MCNs from the LDOS, the LDOS for each unit cell in the has been integrated over the lower band (Figs. \ref{Fig4}(a,b,e,f)), and an in-gap spectra range taking into account all corner modes (Figs. \ref{Fig4}(c,d,g,h)). As displayed, the experimentally measured data agree well with theoretically simulated results.

We see that the measured lower-band LDOS for each bulk unit cell is approximately $\rho=2$, which conforms to the expectation at half filling (the lower band is twofold degenerate). In contrast, due to the corner modes, the lower-band LDOS $\rho^{\text{band}}_c$ in the shaded corner unit cells are no longer 2. As shown, at each corner, for $N=4$ there are four corner unit cells within which the LDOS is closing to $1.5$, while only one for $N=1$. We refer to such unit cell as corner-localized unit cell marked by shaded backgrounds in Fig. \ref{Fig4}. Then the numbers of corner-localized unit cells detect the numbers of corner modes $N$. The fractional corner mode densities $\rho\approx1.5$ also manifests the $\frac{1}{2}$ fractional corner charges. As a result, for the $\mathbb{Z}$-class topological phases $N=4$, there are four $\frac{1}{2}$ fractional corner charges equally distributed at each corner, which is distinct from the $\mathbb{Z}_2$-class topological phases hosting one such fractional charge. The numbers of corner modes $N$ also can be detected by directly looking at the in-gap LDOS that is contributed solely by corner modes, as shown in Figs. \ref{Fig4}(c,d,g,h). As expected, the LDOS for each bulk unit cell is almost zero and only corner LDOS dominate. As the corner modes have weights in the edge and bulk nodes, the in-gap LDOS $\rho^{\text{gap}}_c$ in the shaded corner unit cells is closing to but not 1.
Likewise, the numbers of such corner-localized unit cell give the values of topological number $N$. Given the obvious difference between the LDOS in the corner-localized and the rest of unit cells, such indicator is a robust observable for experimental detection.

In summary, we have experimentally observed $\mathbb{Z}$-class HOTI phases with large-number corner modes in electric circuits. By fabricated different long-range-coupling topolectrical circuits and measuring the impedances in all circuit nodes, we have realized four distinct $\mathbb{Z}$-class HOTI phases characterized by the MCNs $N=1,2,3,4$, and detected the corresponding zero-energy corner-mode resonances and density distributions. Our experimental results, agreeing well with theoretical simulations, clearly demonstrate that the corner modes in $\mathbb{Z}$-class HOTI phases have much larger mode areas at each corner, scaling up with the topological number $N$. Based on LDOS measurements, we have exhibited how to probe the numbers of corner modes $N$, which allows for unambiguous distinguishing different $\mathbb{Z}$-class HOTI phases. With the LDOS, we have also found that for $\mathbb{Z}$-class HOTI phases, there are $N$ fractional corner charges $\frac{1}{2}$ orderly distributed at each corner.

The realization of large topological numbers and multiple corner modes in one corner structure will have several important consequences. Firstly, higher-order topological pump~\cite{HOTIpump,Xie1} has been recently achieved based on $\mathbb{Z}_2$ HOTI phases. However, with our observed $\mathbb{Z}_2$ HOTI phases, four corner states at each corner feature degenerate evolution and may lead to non-abelian topological pump~\cite{Xie2}. Secondly,
creating a disclination structure from our lattice can trap large-number degenerate topological disclination states at the disclination core, thus going beyond previous $\mathbb{Z_2}$ topological disclination states~\cite{Lin1,TL1,Hughes3}. Thirdly, when gain and loss are introduced in our circuits~\cite{Thomale2020,Thomale2021,Hughes2023}, one can study PT symmetry phase transitions~\cite{Christodoulides1,Boettcher1} in our circuit lattices. Due to the four corner states at each corner, the non-Hermitian HOTI phase diagram might be more diverse than the counterpart of $\mathbb{Z}_2$ class. Finally, the large-number corner modes with scalable mode areas can hold promise for enabling exotic topological laser~\cite{Park2020,Xu2020a}.

\emph{Note added}.
We note a very recent work that reports a similar observation of $\mathbb{Z}$-class HOTI phases in acoustic systems~\cite{Jing2023}. Compared to their work, based on LODS measurements, our work has probed the Z-class topological invariants $N$ and revealed the unique fractional corner charges associated with $\mathbb{Z}$-class HOTI phases.

\begin{acknowledgments}
\emph{Acknowledgment}. This work was supported by the National Key Research and Development Program of China (Grant No. 2022YFA1404201), National Natural Science Foundation of China (NSFC) (Grant No. 12034012, 12074234), Stable Support Program for Higher Education Institutions of Shenzhen (No.20220817185604001),
Start-up funding at the Chinese University of Hong Kong, Shenzhen (UDF01002563), Changjiang Scholars and Innovative Research Team in University of Ministry of Education of China (PCSIRT)(IRT\_17R70), Fund for Shanxi 1331 Project Key Subjects Construction, 111 Project (D18001).
\end{acknowledgments}


\begin{thebibliography}{99}


\bibitem{Kane2010}  M.Z. Hasan, and C. L. Kane, Colloquium: topological insulators, Rev. Mod. Phy.\textbf{82},3045 (2010).
\bibitem{Zhang2011} X. L. Qi, and S. C. Zhang, Topological insulators and superconductors, Rev. Mod. Phy. \textbf{83}, 1057 (2011).
\bibitem{Lu2014rev}  L. Lu, J. D. Joannopoulos, and  M. Solja\v{c}i\'{c}, Topological photonics, Nat. photonics \textbf{8}, 821 (2014).
\bibitem{Huber2016rev}  S. D. Huber,  Topological mechanics, Nat. Phys \textbf{12}, 621-623 (2016).
\bibitem{Khan2017rev} A. B. Khanikaev and G. Shvets, Two-dimensional topological photonics, Nat. photonics \textbf{11}, 763 (2017).
\bibitem{Lu2018rev} X. Zhang, M. Xiao, Y. Cheng, M. H. Lu,  and  J. Christensen,  Topological sound, Commun. Phys. \textbf{1}, 97 (2018).
\bibitem{Lu2019rev} T. Ozawa,  H. M. Price, A. Amo, N. Goldman,  M. Hafezi, L. Lu , and  I. Carusotto,  Topological photonics, Rev. Mod. Phy. \textbf{91}, 015006 (2019).
\bibitem{Chan2019rev}  G. Ma, M. Xiao, and C. T. Chan,  Topological phases in acoustic and mechanical systems, Nat. Rev. Phys. \textbf{1}, 281 (2019).
\bibitem{Zhang2022rev} H. Xue, Y. Yang, and B. Zhang, Topological acoustics, Nat. Rev. Mater. \textbf{7}, 974 (2022).
\bibitem{Alu2022rev} X. Ni, S. Yves, A. Krasnok, and A. Alu, Topological Metamaterials, arXiv:2211.10006 (2022).

\bibitem{Wang2013} J. Wang, B. Lian, H. Zhang, Y. Xu, and S. C. Zhang,  Quantum anomalous Hall effect with higher plateaus, Phys. Rev. Lett. \textbf{111}, 136801 (2013).

\bibitem{Fang2014} C. Fang, M. J. Gilbert, and B. A. Bernevig. Large-Chern-number quantum anomalous Hall effect in thin-film topological crystalline insulators, Phys. Rev. Lett. \textbf{112},046801 (2014).

\bibitem{An2016} T.S. Xiong, J. Gong, and J.H. An, Towards large-Chern-number topological phases by periodic quenching, Phys. Rev. B \textbf{93}, 184306 (2016).

\bibitem{Marin2014}  S. A. Skirlo, L. Lu, and  M. Solja\v{c}i\'{c}, Multimode one-way waveguides of large Chern numbers. Phys. Rev. Lett. \textbf{113}, 113904 (2014).

\bibitem{Marin2015} S. A. Skirlo, L. Lu, Y. Igarashi, Q. Yan, J. Joannopoulos, and M. Solja\v{c}i\'{c},  Experimental observation of large Chern numbers in photonic crystals. Phys. Rev. Lett. \textbf{115}, 253901 (2015).

\bibitem{Zhao2020}  Y. F. Zhao, R. Zhang, R. Mei, L.J. Zhou, H. Yi, Y.Q. Zhang, and C.Z. Chang, Tuning the Chern number in quantum anomalous Hall insulators, Nature, \textbf{588}, 419-423 (2020).


\bibitem{HOTI2017a} W. A. Benalcazar, B. A. Bernevig, and T. L. Hughes, Quantized electric multipole insulators, Science \textbf{357}, 61 (2017).

\bibitem{HOTI2017b} J. Langbehn, Y. Peng, L. Trifunovic, F. von Oppen, and P. W. Brouwer, Reflection-Symmetric Second-Order Topological Insulators and Superconductors, Phys. Rev. Lett. \textbf{119}, 246401 (2017).

\bibitem{HOTI2017c} Z. Song, Z. Fang, and  C. Fang,  (d-2)-dimensional edge states of rotation symmetry protected topological states, Phys. Rev. Lett. \textbf{119}, 246402 (2017).

\bibitem{HOTI2018} F. Schindler, A. M. Cook, M. G. Vergniory, Z. Wang, S. S. Parkin, B. A. Bernevig, and T. Neupert, Higherorder topological insulators, Sci.Adv \textbf{4}, eaat0346 (2018).

\bibitem{Xie2021rev} B. Xie, H.-X. Wang, X. Zhang, P. Zhan, J.-H. Jiang, M. Lu, and Y. Chen, Higher-order band topology, Nat. Rev. Phys. \textbf{3}, 520 (2021).

\bibitem{Jacob2020rev} M. Kim, Z. Jacob, and J. Rho, Recent advances in 2D, 3D and higher-order topological photonics, LightSci.Appl \textbf{9}, 130 (2020).

\bibitem{Bena2018a} C. W. Peterson, W. A. Benalcazar, T. L. Hughes, and G. Bahl, A quantized microwave quadrupole insulator with topologically protected corner states Nature, \textbf{555}, 346 (2018).

\bibitem{Bena2018b} J. Noh, W. A. Benalcazar, S. Huang, M. J. Collins, K. P. Chen, T. L. Hughes, and M. C. Rechtsman, Topological protection of photonic mid-gap defect modes, Nat. Photonics \textbf{12}, 408 (2018).

\bibitem{Ashraf2019} A. E. Hassan, F. K. Kunst, A. Moritz, G. Andler, E. J. Bergholtz, M. Bourennane, Corner states of light in photonic waveguides, Nat. Photonics \textbf{13}, 697 (2019).

\bibitem{Lu2019} B. Y. Xie, G. X. Su, H. F. Wang, H. Su, X. P. Shen, P. Zhan, M.-H. Lu, Z. L. Wang, and Y. F. Chen, Visualization of Higher-Order Topological Insulating Phases in Two-Dimensional Dielectric Photonic Crystals, Phys. Rev. Lett. \textbf{122}, 233903 (2019).

\bibitem{Dong2019} X. D. Chen , W. M. Deng, F. L. Shi , F. L. Zhao,  M. Chen, and J. W. Dong,  Direct observation of corner states in second-order topological photonic crystal slabs, Phys. Rev. Lett. \textbf{122}, 233902 (2019).

\bibitem{Hafezi2019} S. Mittal, V. V. Orre, G. Zhu, M. A. Gorlach, A. Poddubny, and M. Hafezi, Photonic quadrupole topological phases, Nat. Photonics \textbf{13}, 692 (2019).

\bibitem{Lu2020} B. Xie, G. Su, H.-F. Wang, F. Liu, L. Hu, S.-Y. Yu, P. Zhan, M.-H. Lu, Z. Wang, and Y.-F. Chen, Higher-order quantum spin Hall effect in a photonic crystal, Nat.Commun. \textbf{11}, 3768 (2020).

\bibitem{Wang2021} Y. Wang,  B. Y. Xie, Y. H. Lu , Y. J. Chang, H. F. Wang, J. Gao, and  X. M. Jin, Quantum superposition demonstrated higher-order topological bound states in the continuum, Light Sci. Appl. \textbf{10}, 173 (2021).

\bibitem{Freqymann2022} J. Schulz, J. Noh, W. A. Benalcazar, G. Bahl, and G. von Freymann, Photonic quadrupole topological insulator using orbital-induced synthetic flux, Nat. Commun. \textbf{13}, 6597 (2022).

\bibitem{Chen2022} Y. Zhang, D. Bongiovanni, Z. Wang, X. Wang, S. Xia, Z. Hu, Z. Chen,  Photonic p-orbital higher-order topological insulators, arXiv preprint arXiv:2208.05961 (2022).

\bibitem{Serra2018} M. S. Garcia, V. Peri, R. Susstrunk, O. R. Bilal, T. Larsen, L. G. Villanueva, and S. D. Huber, Observation of a phononic quadrupole topological insulator, Nature \textbf{555}, 342 (2018).

\bibitem{Xue2019} H. Xue, Y. Yang, F. Gao, Y. Chong, and B. Zhang, Acoustic higher-order topological insulator on a kagome lattice, Nat. Mater \textbf{18}, 108 (2019).

\bibitem{Khan2019} X. Ni, M. Weiner, A. Alu, and A. B. Khanikaev, Observation of higher-order topological acoustic states protected by generalized chiral symmetry, Nat. Mater \textbf{18}, 113 (2019).

\bibitem{Jiang2019a}  X. Zhang, H. X. Wang, Z. K. Lin, Y. Tian,  B. Xie,  M. H. Lu, and J. H. Jiang, Second-order topology and multidimensional topological transitions in sonic crystals, Nat. Phys. \textbf{15}, 582-588 (2019).
\bibitem{Jiang2019b} X. Zhang, B. Y. Xie, H. F. Wang, X. Xu, Y. Tian, J. H. Jiang, M. H. Lu, and Y. F. Chen, Dimensional hierarchy of higher-order topology in three-dimensional sonic crystals, Nat. Commun. \textbf{10}, 5331 (2019).
\bibitem{Zhang2019} H. Xue, Y. Yang, G. Liu, F. Gao, Y. Chong, and B. Zhang, Realization of an Acoustic Third-Order Topological Insulator, Phys. Rev. Lett. \textbf{122}, 244301 (2019).
\bibitem{Liu2020} Y. Qi, C. Qiu, M. Xiao, H. He, M. Ke, and Z. Liu, Acoustic Realization of Quadrupole Topological Insulators, Phys. Rev. Lett. \textbf{124}, 206601 (2020).
\bibitem{Khan2020} X. Ni, M. Li, M. Weiner, A. Alu, and A. B. Khanikaev, Demonstration of a quantized acoustic octupole topological insulator, Nat. Commun. \textbf{11}, 2108 (2020).
\bibitem{Liu2021} Y. Yang, J. Lu, M. Yan, X. Huang, W. Deng, and Z. Liu, Hybrid-Order Topological Insulators in a Phononic Crystal, Phys. Rev. Lett. \textbf{126}, 156801 (2021).
\bibitem{Qiu2021} H. Qiu, M. Xiao, F. Zhang, and C. Qiu, Higher-Order Dirac Sonic Crystals, Phys. Rev. Lett. \textbf{127}, 146601 (2021).
\bibitem{Jiang2022a}  S. Q. Wu,  Z. K. Lin,  B. Jiang, X. Zhou, Z. H. Hang, B. Hou, J. H. Jiang, Higher-order Topological States in Acoustic Twisted Moire Superlattices, Phys. Rev. Appl. \textbf{17}, 034061 (2022).
\bibitem{Jiang2022b} S. Zheng, X. Man, Z. L. Kong, Z. K. Lin, G. Duan, N. Chen, and B. Xia,  Observation of fractal higher-order topological states in acoustic metamaterials, Sci. Bull. \textbf{67}, 2069-2075 (2022).
\bibitem{Junkai2022} J. Li, Q. Mo, J. H. Jiang, and Z. Yang,  Higher-order topological phase in an acoustic fractal lattice, arXiv preprint arXiv:2205.05298 (2022).
\bibitem{Imhof2018}  S. Imhof, C. Berger, F. Bayer,  J. Brehm, L. W. Molenkamp, T. Kiessling, and R. Thomale,  Topolectrical-circuit realization of topological corner modes, Nat. Phys. \textbf{14}, 925-929 (2018).
\bibitem{Ezawa2018} M. Ezawa,  Higher-order topological electric circuits and topological corner resonance on the breathing kagome and pyrochlore lattices, Phys. Rev. B  \textbf{98}, 201402 (2018).
\bibitem{ZhangX2019} J. Bao, D. Zou, W. Zhang, W. He, H. Sun, and  X. Zhang, Topoelectrical circuit octupole insulator with topologically protected corner states, Phys. Rev. B  \textbf{100}, 201406 (2019).
\bibitem{LiuZ2020} J. Wu, X. Huang, J. Lu, Y. Wu, W. Deng, F. Li, and Z. Liu,  Observation of corner states in second-order topological electric circuits, Phys. Rev. B, \textbf{102}, 104109 (2020).
\bibitem{Yan2020a} H. Yang, Z. X. Li,  Y. Liu, Y. Cao, and P. Yan, Observation of symmetry-protected zero modes in topolectrical circuits, Phys. Rev. Res. \textbf{2}, 022028 (2020).
\bibitem{Yan2020b} L. Song, H. Yang, Y. Cao, and P. Yan, Realization of the square-root higher-order topological insulator in electric circuits, Nano Lett. \textbf{20}, 7566-7571 (2020).
\bibitem{ZhangS2020} S. Liu, S. Ma, Q. Zhang, L. Zhang, C. Yang, O. You, and S. Zhang, Octupole corner state in a three-dimensional topological circuit, Light Sci. Appl. \text{9}, 145 (2020).
\bibitem{ZhangX2020} W. Zhang, D. Zou, J. Bao, W. He, Q. Pei, H. Sun, and X. Zhang, Topolectrical-circuit realization of a four-dimensional hexadecapole insulator, Phys. Rev. B  \textbf{102}, 100102 (2020).
\bibitem{ZhangX2021} W. Zhang, D. Zou, Q. Pei, W. He, J. Bao, H. Sun, and X. Zhang, Experimental observation of higher-order topological Anderson insulators, Phys. Rev. Lett. \textbf{126}, 146802 (2021).
\bibitem{Xu2021} B. Lv, R. Chen, R. Li, C. Guan, B. Zhou, G. Dong, and D. H. Xu, Realization of quasicrystalline quadrupole topological insulators in electrical circuits, Commun. Phys. \textbf{4}, 108 (2021).
\bibitem{Hughes2022} S. S. Yamada, T. Li, M. Lin, C. W. Peterson, T. L. Hughes, and G. Bahl, Bound states at partial dislocation defects in multipole higher-order topological insulators, Nat. Commun. \textbf{13}, 2035 (2022).
\bibitem{Gorlach2022} N. A. Olekhno, A. D. Rozenblit, V. I. Kachin, A. A. Dmitriev, O. I. Burmistrov, P. S. Seregin, and M. A. Gorlach, Experimental realization of topological corner states in long-range-coupled electrical circuits, Phys. Rev. B  \textbf{105}, L081107 (2022).
\bibitem{ZhangX2022a}  W. Zhang, H. Yuan, N. Sun, H. Sun, and X. Zhang, Observation of novel topological states in hyperbolic lattices, Nat. Commun. \textbf{13}, 2937 (2022).
\bibitem{Yan2022}  H. Yang, L. Song, Y. Cao, and P. Yan, Observation of type-III corner states induced by long-range interactions, Phys. Rev. B \textbf{106}, 075427 (2022).
\bibitem{ZhangX2022b}  X. Zheng, T. Chen, and X. Zhang, Topolectrical circuit realization of quadrupolar surface semimetals, Phys. Rev. B  \textbf{106}, 035308 (2022).
\bibitem{Park2020}  H. R. Kim, M. S. Hwang, D. Smirnova, K. Y. Jeong,  Y. Kivshar, and H. G. Park, Multipolar lasing modes from topological corner states, Nat. Commun. \textbf{11}, 5758 (2020).
\bibitem{Xu2020a} W. Zhang, X. Xie,  H. Hao, J. Dang, S. Xiao, S. Shi, and X. Xu, Low-threshold topological nanolasers based on the second-order corner state, Light Sci. Appl. \textbf{9}, 109 (2020).
\bibitem{Fleury2019}  F. Zangeneh-Nejad, and R. Fleury, Nonlinear second-order topological insulators, Phys. Rev. Lett. \textbf{123}, 053902 (2019).
\bibitem{Heinrich2021}  M. S. Kirsch, Y. Zhang, M. Kremer, L. J. Maczewsky, S. K. Ivanov, Y. V. Kartashov, and M. Heinrich, Nonlinear second-order photonic topological insulators, Nat. Phys.  \textbf{17}, 995-1000 (2021).
\bibitem{Chen2021}  Z. Hu, D. Bongiovanni, D. Juki\'{c}, E. Jajti\'{c}, S. Xia, D. Song, and Z. Chen, Nonlinear control of photonic higher-order topological bound states in the continuum, Light Sci. Appl. \textbf{10}, 164 (2021).
\bibitem{Ezawa2022} M. Ezawa, Nonlinear non-Hermitian higher-order topological laser, Phys. Rev. Res. \textbf{4}, 013195 (2022).
\bibitem{Xu2020b}  X. Xie, W.Zhang, X. He, S. Wu, J. Dang, K. Peng, and X. Xu, Cavity quantum electrodynamics with second-order topological corner state, Laser Photonics Rev. \textbf{14}, 1900425 (2020).

\bibitem{Li2022} C. Li, M. Li, L.Yan, S. Ye, X. Hu, Q. Gong, and  Y. Li,  Higher-order topological biphoton corner states in two-dimensional photonic lattices, Phys. Rev. Res. \textbf{4}, 023049 (2022).
\bibitem{Bena2022}  W. A. Benalcazar, and  A. Cerjan, Chiral-symmetric higher-order topological phases of matter. Phys. Rev. Lett. \textbf{128}, 127601 (2022).
\bibitem{Lee2018}  C. H. Lee, S. Imhof, C. Berger, F. Bayer, J. Brehm, L. W. Molenkamp, and R. Thomale, Topolectrical circuits, Commun. Phys. \textbf{1}, 39 (2018).
\bibitem{Zhao2018} E. Zhao,  Topological circuits of inductors and capacitors, Ann. Phys. \textbf{399}, 289-313 (2018).
\bibitem{Roy2021} J. Dong, V. Juri\v{c}i\'{c}, and  B. Roy, Topolectric circuits: theory and construction, Phys. Rev. Res. \textbf{3}, 023056 (2021).


\bibitem{HOTIpump} Benalcazar, W. A., Noh, J., M. Wang, S. Huang, K. P. Chen, and  M. C. Rechtsman,  Higher-order topological pumping and its observation in photonic lattices, Phys. Rev. B  \textbf{105}, 195129 (2022).

\bibitem{BahlG2020} C. W. Peterson, T. Li, W. A. Benalcazar, T. L. Hughes, and G. Bahl, A fractional corner anomaly reveals higher-order topology, Science  \textbf{368}, 1114 (2020).

\bibitem{Dis2021a} C. W. Peterson, T. Li, W. Jiang, T. L. Hughes, and G. Bahl, Trapped fractional charges at bulk defects in topological insulators, Nature, \textbf{589}, 376 (2021).

\bibitem{Dis2021b} Y. Liu,  S. Leung, F. F. Li, Z. K. Lin, X. Tao, Y. Poo, and J. H. Jiang, Bulk–disclination correspondence in topological crystalline insulators, Nature, \textbf{589}, 381 (2021).

\bibitem{JiangJH2022} S. Leung, Y. Liu, F. F. Li, C. Liang, Y. Poo, and J. H. Jiang, Observation of fractional quantum numbers at photonic topological edges and corners, arXiv preprint arXiv:2203.00206 (2022).

\bibitem{ZhangS2022b} B. Xie, R. Huang, S. Jia, Z. Lin, J. Hu, Y. Jiang, and  S. Zhang, Bulk-LDOS Correspondence in Topological Insulators, arXiv preprint arXiv:2209.02347  (2022).

\bibitem{GeH2023} H. Ge, Z. W. Long, X. Y. Xu, J. G. Hua, Y. Liu, B. Y. Xie, and Y. F. Chen, Direct measurement of acoustic spectral density and fractional topological charge, Phys. Rev. Appl. \textbf{19}, 034073 (2023).



\bibitem{HOTIpump} Higher-order topological pumping and its observation in photonic lattices
Wladimir A. Benalcazar, Jiho Noh, Mohan Wang, Sheng Huang, Kevin P. Chen, and Mikael C. Rechtsman
Phys. Rev. B 105, 195129 – Published 20 May 2022

\bibitem{Xie1} B. Y. Xie, O. You, and S. Zhang, B. Y. Xie, O. You, and  S. Zhang, Photonic topological pump between chiral disclination states, Phys. Rev. A, \textbf{106}, L021502 (2022).

\bibitem{Xie2} O. You, S. Liang, B. Xie, W. Gao, W. Ye, J. Zhu, and S. Zhang,  Observation of non-Abelian Thouless pump, Phys. Rev. Lett. \textbf{128}, 244302 (2022).



\bibitem{Lin1} A. Rüegg, and C. Lin, Bound states of conical singularities in graphene-based topological insulators, Phys. Rev. Lett. \textbf{110}, 046401 (2013).
\bibitem{TL1} J. C. Teo, and T. L. Hughes, Existence of Majorana-fermion bound states on disclinations and the classification of topological crystalline superconductors in two dimensions, Phys. Rev. Lett. \textbf{111}, 047006 (2013).
\bibitem{Hughes3} T. Li, P. Zhu, W. A. Benalcazar, and T. L. Hughes, Fractional disclination charge in two-dimensional C n-symmetric topological crystalline insulators, Phys. Rev. B \textbf{101}, 115115 (2020).

\bibitem{Thomale2020} T. Helbig, T. Hofmann, S. Imhof, M. Abdelghany, T. Kiessling, L. W. Molenkamp, and  R. Thomale, Generalized bulk–boundary correspondence in non-Hermitian topolectrical circuits. Nat. Phys.  \textbf{16}, 747-750 (2020).
\bibitem{Thomale2021} A. Stegmaier, S. Imhof, T. Helbig, T. Hofmann, C. H. Lee, M. Kremer, and Thomale, R.  Topological defect engineering and P T symmetry in non-hermitian electrical circuits, Phys. Rev. Lett. \textbf{126}, 215302 (2021).
\bibitem{Hughes2023} P. Zhu, X. Q. Sun, T. L. Hughes, and  G. Bahl,  Higher rank chirality and non-Hermitian skin effect in a topolectrical circuit, Nat. Commun. \textbf{14}, 720 (2023).

\bibitem{Christodoulides1} R. El-Ganainy, K. G. Makris, M. Khajavikhan, Z. H. Musslimani, S. Rotter, and D. N. Christodoulides, Non-Hermitian physics and PT symmetry, Nat. Phys. \textbf{14}, 11-19 (2018).
\bibitem{Boettcher1} C. M. Bender, and S. Boettcher, Real spectra in non-Hermitian Hamiltonians having PT symmetry, Phys. Rev. Lett. \textbf{80}, 5243 (1998).

\bibitem{Jing2023} D. Wang, Y. Deng, M. Oudich, W. A. Benalcazar, G. Ma, Y. Jing, Realization of a Z-classified chiral-symmetric higher-order topological insulator in a coupling-inverted acoustic crystal, arXiv:2305.08313.





\end{thebibliography}
\end{document}